\documentclass[reprint,aps,pra,showpacs,superscriptaddress]{revtex4-1} 
\usepackage{amsmath,amssymb}
\usepackage{mathrsfs}
\usepackage{bbm}
\usepackage[dvipdfmx]{graphicx, color}
\usepackage{dcolumn}
\usepackage{bm}
\usepackage{enumerate}
\usepackage{theorem}
\newtheorem{theorem}{Theorem}
\newtheorem{lemma}{Lemma}

\def\argmin{\mathop{\mathrm{argmin}}}
\def\argmax{\mathop{\mathrm{argmax}}}
\def\trace{\mathop{\mathrm{tr}}}
\def\Trace{\mathop{\mathrm{Tr}}}
\def\est{\mathop{\mathrm{est}}}
\def\Aopt{\mathop{\mathrm{A\mbox{\scriptsize -}opt}}}

\begin{document}
\title{Adaptive experimental design for one-qubit state estimation with finite data based on a statistical update criterion}



\author{Takanori Sugiyama}
\email{sugiyama@eve.phys.s.u-tokyo.ac.jp}
\affiliation{
Department of Physics, Graduate School of Science, The University of Tokyo,\\
7-3-1 Hongo, Bunkyo-ku, Tokyo, Japan 113-0033.
}
\author{Peter S. Turner}
\email{turner@phys.s.u-tokyo.ac.jp}
\affiliation{
Department of Physics, Graduate School of Science, The University of Tokyo,\\
7-3-1 Hongo, Bunkyo-ku, Tokyo, Japan 113-0033.
}
\author{Mio Murao}
\email{murao@phys.s.u-tokyo.ac.jp}
\affiliation{
Department of Physics, Graduate School of Science, The University of Tokyo,\\
7-3-1 Hongo, Bunkyo-ku, Tokyo, Japan 113-0033.
}
\affiliation{
Institute for Nano Quantum Information Electronics, The University of Tokyo,\\
4-6-1 Komaba, Meguro-ku, Tokyo, Japan 153-8505.
}
\date{\today}
\begin{abstract}
We consider 1-qubit mixed quantum state estimation by adaptively updating measurements according to previously obtained outcomes and measurement settings.
Updates are determined by the average-variance-optimality (A-optimality) criterion, known in the classical theory of experimental design and applied here to quantum state estimation.
In general, A-optimization is a nonlinear minimization problem; however, we find an analytic solution for 1-qubit state estimation using projective measurements, reducing computational effort.
We compare numerically two adaptive and two nonadaptive schemes for finite data sets and show that the A-optimality criterion gives more precise estimates than standard quantum tomography.
\end{abstract}
\maketitle

%
\section{Introduction}
For successful experimental implementation of any quantum protocol, the quantum states and operations involved must be confirmed to be sufficiently closed to their theoretical targets.
One way to obtain such a confirmation is to perform another experiment and from the obtained data make an estimate of the quantum operator involved.
Statistically, this is a constrained multi-parameter estimation problem -- the quantum estimation problem -- where we assume we are given a finite number of identical copies of a quantum state or operation, we perform measurements whose mathematical description is assumed to be known, and from the outcome statistics we make our estimate.
Due to the probabilistic behavior of the measurement outcomes and the finiteness of the number of measurement trials, there always exist statistical errors in any quantum estimate. 
The size of the error depends on the choice of measurements and the estimation procedure.
In statistics, the former is called an \emph{experimental design}, while the latter is called an \emph{estimator}. 
It is, therefore, a key aim of both classical and quantum estimation theory to find a combination of experimental design and estimator which gives us more precise estimation results using fewer measurement trials.

A standard combination in quantum information experiments is that of quantum tomography and maximum likelihood estimator.  
Although the term ``quantum tomography'' can be used in several different contexts, we use it to mean an experimental design in which an independently and identically prepared set of measurements are used throughout the entire experiment \cite{QTText2004}.
The performance of different choices for the set of tomographic measurements have been studied, in, for example, \cite{Burgh2008, Nunn2010}. 
This of course raises the question of the performance of {\it adaptive} experimental designs, in which the measurements performed from trial to trial are not independent, and are chosen according to previous measurement settings and the outcomes obtained.
Clearly, adaptive experimental designs are a superset of the nonadaptive ones, and as such can potentially achieve higher performance.

Adaptive designs are characterized by the way in which measurements are related from trial to trial, referred to as an \emph{update criterion}.
Previously proposed update criteria include those based on asymptotic statistical estimation theory (Fisher information) \cite{Nagaoka1988, NagaokaChap10, Fujiwara2006},  direct calculations of the estimates expected to be obtained in the next measurement \cite{Fischer2000, Happ2008}, mutually unbiased basis \cite{HappFreyberger2011}, as well as Bayesian estimators and Shannon entropy \cite{Fischer2000, FischerFreyberger2000_PLA, Huszar2011}. 
Theoretical investigations report that some of the proposed update criteria give more precise estimates than nonadaptive quantum tomography, and an experimental implementation of the update criterion proposed in \cite{Fischer2000} in an ion trap system has been performed \cite{Hannemann2002}.
If $N$ denotes the number of measurement trials and $N$ is sufficiently large, it is known in 1-qubit state estimation that the expectation value of infidelity averaged over states, a measure of the estimation error, can decrease at best as $O(N^{-3/4})$ in a nonadaptive experiment \cite{Bagan2004}, compared to $O(N^{-1})$ in adaptive experiments \cite{Bagan2006PRL}.
Most of the proposed update criteria, however, have high computational cost that makes real experiments infeasible.
In this paper, we propose an adaptive experimental design whose average expected infidelity decreases as $O(N^{-1})$ and whose update criterion, known as average-variance optimality (A-optimality) in classical statistics, has low computational cost  for 1-qubit state estimation.

The paper is structured as follows.
In Sec. \ref{sec:Preliminaries} we lay out the notation and terminology that will be used throughout in this paper, by explaining basic concepts in adaptive experimental design, statistical parameter estimation, and A-optimality criteria. 
We also give a brief review of some of the proposed update criteria in the literature.
In Sec. \ref{section:Result} we give the explicit form of the analytic solution of the A-optimal update criterion, (the derivation is given in the Appendix).
This analytic solution makes it possible to reduce the computational cost for updating measurements, and using this we compare several estimation schemes numerically, showing that our proposal is more precise than standard quantum tomography. 
In Sec. \ref{section:discussion}, we discuss the feasibility of implementing the proposed scheme experimentally. 
A summary appears in Sec. \ref{section:summary}.

\section{Preliminaries\label{sec:Preliminaries}}

  \subsection{Notation and terminology}

We will adopt terms from the statistical literature, since they afford us the precision we need to properly discuss details of estimation schemes that can sometimes be subtle.  
In this subsection we will introduce a formalism for quantum estimation using that terminology, and apply it in a survey of several existing update criteria in Sec. \ref{subsec:Survey}. 

  \subsubsection{Model selection}
   
  In statistical estimation theory, a statistical model is defined as a set of probability distributions, and we assume that the true probability distribution of interest is included in the set.
  In the quantum case, a probability distribution is determined by the state of the system and the action of the measurement on the state system.
  Let $\mathcal{H}$ be a Hilbert space with finite dimension $d$ and $\mathcal{S}(\mathcal{H})$ be the set of all density matrices acting on that Hilbert space.
  Suppose we know that the object we are trying to estimate lies in a subset $\mathcal{O} \subseteq \mathcal{S}(\mathcal{H})$, that is, the true density matrix $\rho$ is included in $\mathcal{O}$.  For example, when we know that the true state is pure, $\mathcal{O}$ is the set of all pure states. 
  In this paper, we consider mixed state estimation, and we assume that in our finite $N$ measurement trials we prepare identical copies of an unknown state $\rho \in \mathcal{O}$. 

   \subsubsection{Experimental design} 
    A probability distribution of outcomes in quantum measurement requires not only a density matrix, but also a positive operator valued measure (POVM), $\bm{\Pi}=\{ \Pi_{x}\}_{x\in \mathcal{X}}$, where $\mathcal{X}$ is the set of outcomes. 
    When the measurement is characterized by a POVM $\bm{\Pi}$ and the measured quantum state is characterized by a density matrix $\rho$, the probability distribution of the outcomes is given by Born's rule $p(x; \bm{\Pi}|\rho) = \mbox{Tr}[\Pi_{x} \rho]$, where $\Trace$ denotes the trace operation with respect to $\mathcal{H}$,  (note that in the next subsection, a different trace operation represented as $\trace$, is introduced).
    
   We consider sequential measurements, as opposed to collective measurements, on copies of $\rho$.  
   We will index measurement trials using subscripts $n \in \{ 1,2, \ldots , N\}$, and sequences using superscripts.  
   Thus, for some symbol $A$, $A_n$ is its value taken at the $n$-th trial, while $A^n$ is the sequence $\{ A_1, A_2, \ldots, A_n\}$.  
   We will also try to use calligraphic fonts for supersets.
   Adaptivity in our sense means that the POVM performed at $(n+1)$-th trial can depend on all the previous $n$ trials' outcomes and POVMs.
   
   The measurement class $\mathcal{M}_{n}$ is the set of POVMs which are available at the $n$-th trial. 
   We choose the $n$-th POVM, $\bm{\Pi}_{n} = \{ \Pi_{n, x} \}_{x\in\mathcal{X}_{n}}$ from $\mathcal{M}_{n}$, where $\mathcal{X}_{n}$ denotes the set of measurement outcomes for the $n$-th trial.
   When it is independent of the trial, as is usually the case, we omit the index, using $\mathcal{M}$ for the measurement class and $\mathcal{X}$ for the outcome set.
    Let $x^n = \{ x_{1}, \ldots , x_{n}\}$ denote the sequence of outcomes obtained up to the $n$-th trial, where $x_{i} \in \mathcal{X}_{i}$.
   We will denote the pair of measurement performed and outcome obtained by $D_n=( \bm{\Pi}_{n}, x_{n} ) \in \mathcal{D}_{n}:=\mathcal{M}_{n} \times \mathcal{X}_{n}$, and refer to it as the data for trial $n$.
   The sequence of data up to trial $n$ is thus $D^n = \{ D_1, \ldots , D_n \} \in \mathcal{D}^{n}:=\times_{i=1}^{n}\mathcal{D}_{i}$.
   After the $n$-th measurement, we choose the next, $(n+1)$-th, POVM $\bm{\Pi}_{n+1} = \{ \Pi_{n+1, x}\}_{x\in \mathcal{X}_{n+1}}$ according to the previously obtained data.
    Let $u_{n}$ denote the map from the data to the next measurement, that is, $u_{n}: \mathcal{D}^{n-1} \to \mathcal{M}_{n}$, $\bm{\Pi}_{n} = u_{n}(D^{n-1})$.
    We call $u_{n}$ the measurement update criterion for the $n$-th trial and $u^N := \{ u_{1}, u_2 , \ldots , u_{N} \}$ the measurement update rule.
    Note that $u_1$ is a map from $\emptyset$ to $\mathcal{M}_{1}$ and corresponds to the choice of the first measurement.  

   \subsubsection{Estimator}
   An estimator $\rho^{\mathrm{est}}=\{ \rho^{\mathrm{est}}_{1}, \ldots, \rho^{\mathrm{est}}_{N}\}$ is a set of maps from the data to the model space, $\rho^{\mathrm{est}}_{n}: \mathcal{D}^n \to \mathcal{O}$ so that $\rho^{\mathrm{est}}_{n}(D^{n})\in\mathcal{O}$.
   The estimated density matrix $\rho^{\mathrm{est}}_{n}(D^{n})$ is called the $n$-th estimate.
   We will often omit the data dependency.
   In this paper we use a maximum likelihood estimator $\rho^{\mathrm{ML}}$ defined as 
   \begin{eqnarray}
      \rho^{\mathrm{ML}}_{n} &:=& \argmax_{\sigma \in \mathcal{O}} p(D^n | \sigma), 
   \end{eqnarray}
   where
   \begin{eqnarray}
      p(D^n |\sigma) &:=& \Trace [\Pi_{1 , x_1} \otimes \Pi_{2, x_{2}} \otimes \cdots \otimes \Pi_{n, x_n} \sigma^{\otimes n} ].      
   \end{eqnarray}
   
\begin{figure}[h!]
       \includegraphics[width =0.95\linewidth]{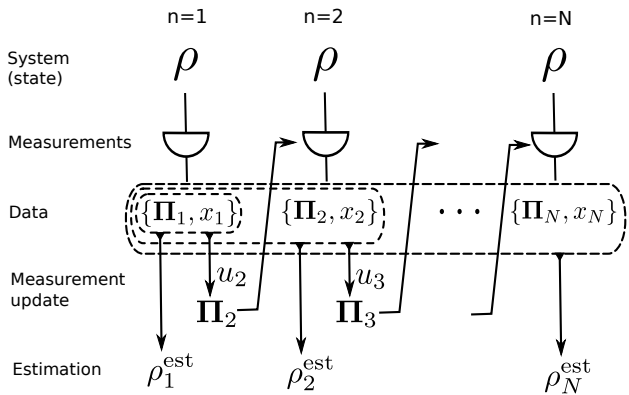}%
       \caption{\label{fig:Sketch} A sketch of a generic procedure for an adaptive quantum estimation scheme.}
   \end{figure}

   A quintuplet $(\mathcal{O}, N, \mathcal{M}^{N}, u^{N}, \rho^{\mathrm{est}})$ specifies an estimation scheme.
   A sketch of the procedure for a generic adaptive quantum estimation scheme is given in Fig. \ref{fig:Sketch}.
   
  \subsubsection{Evaluation}
   In order to evaluate the precision of estimates of the true density matrix, we introduce a \emph{loss function} (sometimes called a cost function).
   A loss function $\Delta$ is a map from $\mathcal{O}\times \mathcal{O}$ to $\mathbb{R}$ such that (i) $\forall \rho, \sigma\in \mathcal{O}, \Delta (\rho ,\sigma) \ge 0$ and (ii) $\forall \rho \in \mathcal{O}, \Delta (\rho , \rho ) = 0$.
   For example, the trace-distance and the infidelity (one minus the fidelity) are loss functions for density matrices.
   The outcomes of quantum measurements are random variables, and the value of the loss function between an estimate and the true density matrix is also a random variable.
   Thus, in order to evaluate the precision of the estimator (not the estimate) for the true density matrix, we use the statistical expectation value of the loss function, called an \emph{expected loss} (sometimes called a risk function) \cite{Note1}.
   The explicit form is given by
   \begin{eqnarray}
     \bar{\Delta}_{N}(u^{N},\rho^{\mathrm{est}}| \rho)&:=&\sum_{D^N \in \mathcal{D}^N} p(D^N | \rho) \Delta (\rho^{\mathrm{est}}_{N}(D^{N}), \rho).     
   \end{eqnarray}
   The value of the expected loss depends on the choice of the estimator as well as the true density matrix.
   The latter is of course unknown in an experiment, and there are at least two approaches to eliminate its dependence, namely the average and the maximal (or worst case) expected loss, given explicitly by
   \begin{eqnarray}
      \bar{\Delta}^{\mathrm{ave}}_{N}(u^{N}, \rho^{\mathrm{est}}) &:=& \int_{\rho \in \mathcal{O}} \!\!\!\!\!\!\! d\mu(\rho) \bar{\Delta}_{N}(u^{N}, \rho^{\mathrm{est}}| \rho), \label{eq:ave}\\
      \bar{\Delta}^{\mathrm{max}}_{N}(u^{N}, \rho^{\mathrm{est}}) &:=& \max_{\rho \in \mathcal{O}} \bar{\Delta}_{N}(u^{N}, \rho^{\mathrm{est}}| \rho). \label{eq:maximalloss}
   \end{eqnarray}
   where $\mu$ is a probability measure on $\mathcal{O}$. 
   The task in this paper is to find a combination of a measurement update rule $u^N$ and estimator $\rho^{\mathrm{est}}$ with average expected loss as small as possible.
       
      
  \subsection{A generalized Cram\'{e}r-Rao inequality} 
  
  The A-optimality criterion is a measurement update criterion based on the asymptotic theory of statistical parameter estimation \cite{LearningText2005, OptimalDesignOfExperiments}.
   In this subsection we introduce a few basic results of the asymptotic theory. 
  First let us parametrize the state space $\mathcal{S}(\mathcal{H})$. 
  Any density matrix on $d$-dimensional Hilbert space can be parametrized by $d^2 -1$ real numbers, $\bm{s} \in \mathbb{R}^{d^2 -1}$, i.e. $\rho = \rho(\bm{s})$. 
  In the $d=2$ case, we take $\rho (\bm{s})= \frac{1}{2}(\mathbbm{1} + \bm{s}\cdot \bm{\sigma}) $, where $\bm{\sigma} = (\sigma_{1}, \sigma_{2}, \sigma_{3})$, $\sigma_{\alpha}\ (\alpha = 1,2,3)$ are the Pauli matrices, and $\bm{s} \in \mathbb{R}^3,\ \| \bm{s} \| \le 1$, is called the Bloch vector. 
  The estimation of $\rho$ is equivalent to the estimation of $\bm{s}$, and we let $\bm{s}^{\mathrm{est}}$ denote the estimator.  
  Estimates of a density matrix and of a Bloch vector are related as $\rho^{\mathrm{est}}_{n}(D^n) = \rho(\bm{s}^{\mathrm{est}}_{n}(D^n))$.

  For any estimator $\bm{s}^{\mathrm{est}}$, any number of measurement trials $N$, and any positive semidefinite matrix $H(\bm{s})$, the inequality
  \begin{align}
     \sum_{D^N \in \mathcal{D}^N} p(D^N |\bm{s}) [ \bm{s}^{\mathrm{est}}_{N}(D^N) - \bm{s} ]^T H(\bm{s}) [ \bm{s}^{\mathrm{est}}_{N}(D^N) - \bm{s} ] \notag \\  
     \ge \trace [H(\bm{s}) G_{N}(u^{N}, \bm{s}^{\mathrm{est}}, \bm{s})^{T} F_{N}(u^{N}, \bm{s})^{-1} G_{N}(u^{N}, \bm{s}^{\mathrm{est}}, \bm{s})] \label{eq:gCRineq}
  \end{align}
  holds, where
  \begin{align}
     p(D^N |\bm{s}) & : = p(D^N | \rho (\bm{s})), \\
     G_{N}(u^{N}, \bm{s}^{\mathrm{est}}, \bm{s}) & := \nabla_{\bm{s}} \sum_{D^N \in \mathcal{D}^N} p(D^N |\bm{s}) \bm{s}^{\mathrm{est} T}_{N}(D^{N}), \\
     F_N (u^{N}, \bm{s}) & := \sum_{D^N \in \mathcal{D}^N} \frac{\nabla_{\bm{s}} p(D^N |\bm{s}) \nabla_{\bm{s}}^{T} p(D^N |\bm{s})}{p(D^N |\bm{s})}, \label{eq: Fisher}
  \end{align}
  and $\trace$ denotes the trace operation with respect to the parameter space.
  Eq.(\ref{eq:gCRineq}) is a known generalization of the Cram\'{e}r-Rao inequality \cite{RaoText1973}, and we give a simple proof in Appendix \ref{sec:ProofCRineq}. 
  $F_N (\bm{s})$ is a $(d^2 -1) \times (d^2 -1)$ positive semidefinite matrix called the Fisher matrix of the probability distribution $\{ p(D^N |\bm{s}) \}_{D^N \in \mathcal{D}^N}$.   
  
  If the estimate converges to the true parameter, i.e., $\bm{s}^{\mathrm{est}}_{N}(D^N) \to \bm{s}$ as $N \to \infty$ with probability 1, the LHS of Eq.(\ref{eq:gCRineq}) converges to 0 and therefore the RHS should converge to 0. 
  In this case, if we assume the exchangeability of the limit and derivative, the matrix $G_{N}(u^{N}, \bm{s}^{\mathrm{est}}, \bm{s})$ converges to the identity matrix $I$, and the quantity $K_{N}(u^{N}, \bm{s})$ defined as
   \begin{eqnarray}
      K_{N}(u^{N}, \bm{s}) := \trace [H(\bm{s}) F_{N}(u^{N}, \bm{s})^{-1}] 
   \end{eqnarray}
   converges to $0$.
   This $K_{N}(u^{N}, \bm{s})$ can be interpreted as a lower bound of the weighted (by $H(\bm{s})$) mean squared error when $N$ is sufficiently large.
    It is known that under certain regularity conditions, a maximum likelihood estimator achieves the equality of Eq.(\ref{eq:gCRineq}) asymptotically.
    For a given $\bm{s}$, it would be wise to choose a measurement update rule which makes the value of $K_{N}(u^{N}, \bm{s})$ as small as possible.   
    This is the guiding principle of the A-optimality criterion. 
  
  \subsection{A-optimality criteria}
  We move on to the explanation of the procedure of A-optimality.
  The ``A" stands for ``average-variance" \cite{OptimalDesignOfExperiments}.
  According to the asymptotic theory of statistical parameter estimation described in the previous subsection, we wish to minimize the value of $K_{N}(u^{N}, \bm{s})$.
  Suppose that we perform $n$ trials and obtained the data sequence $D^n$.
  We would like to choose the POVM minimizing $K_{n+1}(u^{N}, \bm{s})$ in $\mathcal{M}_{n+1}$ as the next, $(n+1)$-th, measurement.  
  When we consider minimizing this function, there are two problems.
  In order to avoid them, we introduce two approximations.
  The first problem is that the minimized function depends on the true parameter $\bm{s}$.
  Of course the true parameter is unknown in parameter estimation problems, and we must use an estimate in the update criterion, $\hat{\bm{s}}^{\mathrm{est}}_{n}(D^n)$, instead.
  The mesurement update estimator $\hat{\bm{s}}^{\mathrm{est}}$ is not necessarily the same as $\bm{s}^{\mathrm{est}}$.
  The second problem is that unlike the independent and identically distributed (i.i.d.) measurement case, calculation of the Fisher matrix in the adaptive case requires summing over an exponential amount of data, and is computationally intensive.
  To avoid this problem, we approximate the sum over all possible measurements by that over only those measurements that have been performed:
  \begin{eqnarray}
     F_{n+1}(u^{n+1} , \bm{s}) 
     & \approx \tilde{F}_{n+1}(u^{n+1} , \bm{s}| D^n ) :=\sum_{i=1}^{n+1} F(\bm{\Pi}_{i}, \bm{s} ), \label{eq:AoptApprox2}
  \end{eqnarray}
  where
  \begin{eqnarray}
     F(\bm{\Pi}_{i}, \bm{s}) & := \sum_{x_{i}\in \mathcal{X}_{i}} \frac{ \nabla_{\bm{s}} p(x_{i};\bm{\Pi}_{i}| \bm{s})  \nabla^T_{\bm{s}} p(x_{i}; \bm{\Pi}_{i}| \bm{s}) } {p(x_{i}; \bm{\Pi}_{i}| \bm{s})}, \\
     \bm{\Pi}_{i} & = u_{i} (D^{i-1}),\ i = 1, \cdots , n+1 .
  \end{eqnarray}
  The matrix $F(\bm{\Pi}_{i}, \bm{s})$ is the Fisher matrix for the $i$-th measurement probability distribution $\{ p(x_{i};\bm{\Pi}_{i}| \bm{s} ) \}_{x_{i}\in \mathcal{X}_{i}}$, and $\tilde{F}_{n+1}(u^{n+1}, \bm{s}| D^n )$ is the sum of the Fisher matrices from the first to the ($n+1$)-th trial. 
   Instead of minimizing $K_{n+1}(u^{n+1}, \bm{s})$, we consider the minimization of 
   \begin{eqnarray}
      \tilde{K}_{n+1}(u^{n+1}, \bm{s} | D^{n}) := \trace [ H(\bm{s}) \tilde{F}_{n+1}(u^{n+1}, \bm{s} | D^{n})^{-1} ].
   \end{eqnarray}
   It is known that the convergence of $\tilde{K}_{N}(u^{N}, \bm{s} | D^{N})$ to $0$ is part of a sufficient condition for the convergence of a maximum likelihood estimator \cite{HallHeyde1980}, and this justifies the use of this second approximation.
   We explain the relationship between the conditional and unconditional Fisher matrices with respect to the estimator's convergence in Appendix \ref{sec:ConditionalFisherMatrices}.
   After making these two approximations, we define the A-optimality criterion as
   \begin{align}
      \bm{\Pi}_{n+1}^{\mathrm{A\mbox{\scriptsize -}opt}} 
      &:= u_{n+1}^{\mathrm{A\mbox{\scriptsize -}opt}}(D^{n}) \notag \\
      &= \argmin_{\bm{\Pi}_{n+1} \in \mathcal{M}_{n+1}} \trace [H ( \hat{\bm{s}}^{\est }_{n} ) \tilde{F}_{n+1}(u^{n+1}, \hat{\bm{s}}^{\est}_{n} | D^n )^{-1} ].\label{eq:DefAopt}
   \end{align}
   Finding $\bm{\Pi}_{n+1}^{\mathrm{A\mbox{\scriptsize -}opt}}$ is a nonlinear minimization problem with high computational cost in general.
   In this paper, we derive the analytic solution of Eq.~(\ref{eq:DefAopt}) in the 1-qubit case, reducing the computational cost significantly.
 
  \subsection{Estimation setting \label{subsec:EstimationSetting}}
  
  We consider a 1-qubit mixed state estimation problem, so that $\mathcal{O} = \mathcal{S}(\mathbb{C}^2)$.
  We identify the Bloch parameter space $\{ \bm{s} \in \mathbb{R}^3 | \| \bm{s} \| <1 \}$ with $\mathcal{O}$, where we restrict the true state space to be strictly the interior in order to avoid the possible divergence of the Fisher matrix.
  Suppose that we can choose any rank-1 projective measurement in each trial.
  Let $\bm{\Pi} (\bm{a}) = \{ \Pi_{x}(\bm{a}) \}_{x=\pm}$ denote the POVM corresponding to the projective measurement onto the $\bm{a}$-axis $(\bm{a} \in \mathbb{R}^3 , \| \bm{a} \|=1)$, whose elements can be represented as
  \begin{align}
     \Pi_{\pm}(\bm{a}) = \frac{1}{2}(\mathbbm{1} \pm \bm{a}\cdot \bm{\sigma}).
  \end{align}
  This is the Bloch parametrization of projective measurements.
  We identify the set of parameters $\mathcal{A}=\{ \bm{a} \in \mathbb{R}^3 |\ \| \bm{a} \|=1 \}$ with the measurement class $\mathcal{M}=\{ \mbox{All rank-1 projective measurements on a 1-qubit system} \}$.
  
  For our loss functions, we use both the squared Hilbert-Schmidt distance $\Delta^{\mathrm{HS}}$ and the infidelity $\Delta^{\mathrm{IF}}$ \cite{Bagan2004}:
  \begin{align}
     \Delta^{\mathrm{HS}} (\bm{s}, \bm{s}^{\prime}) 
     :&= \frac{1}{2} \Trace [ \bigl( \rho (\bm{s}) - \rho (\bm{s}^{\prime}) \bigr)^2 ] \\
      &= \frac{1}{4} (\bm{s} - \bm{s}^{\prime} )^2 , \label{eq:HSdistance2_para}\\
     \Delta^{\mathrm{IF}} (\bm{s}, \bm{s}^{\prime} ) 
     :&= 1 - \Trace \bigl[ \sqrt{ \sqrt{\rho (\bm{s})} \rho (\bm{s}^{\prime}) \sqrt{\rho (\bm{s})} } \bigr]^2 \\
      &= \frac{1}{2}\bigl( 1 - \bm{s}\cdot \bm{s}^{\prime} - \sqrt{1-\|\bm{s}\|^2}\sqrt{1-\|\bm{s}^{\prime}\|^2 }  \bigr).\label{eq:Infidelity_para}
  \end{align}  
  We note that the Hilbert-Schmidt distance coincides with the trace distance in a 1-qubit system.
 The asymptotic behavior of the average expected fidelity $\bar{\Delta}^{\mathrm{IF} \mathrm{ave}}_{N}$ is known in the 1-qubit state estimation case \cite{Bagan2004, Bagan2006PRA, Bagan2006PRL}.
 The measure used for calculating this average is the Bures distribution, $d\mu(\bm{s}) = \frac{1}{\pi^2} (1- \| \bm{s} \|^2 )^{-1/2} d\bm{s}$.  
  If we limit our available measurements to be sequential and independent (i.e., nonadaptive), $\bar{\Delta}^{\mathrm{IF} \mathrm{ave}}_{N}$ behaves at best as $O(N^{-3/4})$ \cite{Bagan2004, Bagan2006PRA}.
  On the other hand, if we are allowed to use adaptive, separable, or collective measurements, $\bar{\Delta}^{\mathrm{IF} \mathrm{ave}}_{N} $ can behave as $O(N^{-1})$ \cite{Bagan2006PRL}.  
  In \cite{Bagan2004, Bagan2006PRA, Bagan2006PRL}, the coefficient of the dominant term in the asymptotic limit is also derived.
  
  In Sec. \ref{subsubsec:AveragedMeanLosses}, we show numerical results.
  A maximum likelihood estimator is used, and it is shown that the average expected infidelity of an A-optimal scheme behaves as $O(N^{-1})$, illustrating that the A-optimality criterion is indeed making use of adaptation to outperform nonadaptive schemes.

  \subsection{Survey of some other update criteria\label{subsec:Survey}}
 
  We briefly review some of the other adaptive measurement update criteria proposed in the literature, using our terminology and notation introduced in the previous subsections.
  
       \subsubsection{Two-step adaptation criterion}
     
     Before explaining update criteria that are performed at each and every trial, such as A-optimality, we briefly review a simpler update criterion.
     The two-step adaptation criterion requires the measurement update only once during a measurement sequence.
     We have
     \begin{eqnarray}
       u_{n+1} (D^{n})= 
       \left\{
          \begin{array}{lr}
             \bm{\Pi}_{\mathrm{1st}} & \mbox{if}\ n < N_{\mathrm{1st}} \\
             \bm{\Pi}_{\mathrm{2nd}} & \mbox{if}\ n \ge N_{\mathrm{1st}}
          \end{array}.
          \right.
     \end{eqnarray}
     Thus, for all trials up to and including trial $N_{\mathrm{1st}}$ a fixed POVM $\bm{\Pi}_{\mathrm{1st}}$ is performed, and an estimate is calculated from the obtained data.
     Using that data we choose a new POVM $\bm{\Pi}_{\mathrm{2nd}}$ for the remaining $N_{\mathrm{2nd}}(=N - N_{\mathrm{1st}})$ copies.
     In \cite{HayashiMatsumoto98_Kokyuroku, HayashiMatsumotoChap13,GillMassar2000,Bagan2006PRL}, two-step adaptation criteria are used to prove mathematically an asymptotic bound for weighted mean squared errors in 1-qubit state estimation.  
     In \cite{RehacekEnglertKaszlikowski04, PetzHangosRuppert07}, some numerical results are shown for a few two-step adaptation schemes.
    
    \subsubsection{N88 criterion}
  
    In \cite{Nagaoka1988, NagaokaChap10, Fujiwara2006}, an update criterion based on the Cram\'{e}r-Rao inequality is proposed.
    The update criterion is given by
    \begin{align}
       u_{n+1}(D^{n}) &= \argmin_{\bm{\Pi}\in\mathcal{M}_{n+1}} \trace [H(\hat{\bm{s}}^{\est}_{n}) F(\bm{\Pi}, \hat{\bm{s}}^{\est}_{n})^{-1}].\label{eq:NagaokaUpdate}
    \end{align}
    The difference from the A-optimality criterion is that in Eq. (\ref{eq:NagaokaUpdate}) the Fisher information matrix used in the update does not take into account all $n+1$ measurements, but about only the $(n+1)$-th measurement.
    The advantage of course is that this reduces the computational cost of updates.
    The disadvantage is that when $\mathcal{M}_{n} \: (n=1,2,\ldots )$ consists of informationally incomplete POVMs, as is the case in most experiments, the estimates cannot converge to the true state.
    As explained in Sec. \ref{subsec:EstimationSetting}, in this paper $\mathcal{M}_{n}$ is restricted to rank-1 projective measurements, and in this setting Eq. (\ref{eq:NagaokaUpdate}) does not work well. 
    
    \subsubsection{FKF00 criteria}  
     
     In \cite{Fischer2000}, two update criteria are proposed.
     \begin{enumerate}[(i)]
     \item The first criterion is based on the Shannon entropy of the estimated measurement probability distribution, and is given by
     \begin{align}
        u_{n+1}&(D^{n}) = \argmax_{\bm{\Pi}\in \mathcal{M}_{n+1}}  \notag \\
          &\Bigl(
             -\sum_{x\in\mathcal{X}_{n+1}} p(x; \bm{\Pi}| \hat{\rho}^{\est}_{n}(D^{n})) \ln p(x;  \bm{\Pi}| \hat{\rho}^{\est}_{n}(D^{n}))
          \Bigr).
     \end{align}
     
     \item The second criterion uses a third state estimator $\hat{\hat{\rho}}^{\est}$ such that
     \begin{align}
        \big(u_{n+1}&(D^{n}\big), \hat{\hat{\rho}}^{\est}_{n}(D^{n})) = \argmax_{(\bm{\Pi}, \sigma ) \in \mathcal{M}_{n+1}\times \mathcal{O}} \notag \\
           &\Bigl(
              \sum_{x\in\mathcal{X}_{n+1}} p(x;  \bm{\Pi}| \hat{\rho}^{\est}_{n}(D^{n})) \Delta(\hat{\rho}^{\est}_{n+1}(D^{n+1}), \sigma)
           \Bigr). \label{eq:est3}
     \end{align}  
     \end{enumerate}
     Numerical simulation is performed for the case where $\mathcal{O}$ is the set of 1-qubit pure states and $\mathcal{M}_n$ is the set of projective measurements, while $\rho^{\est}$ is a biased maximum likelihood estimator, $\hat{\rho}^{\est}$ is a Bayesian estimator up to $N=60$.  Average (not expected) infidelity is used as the evaluation function.  
           
     \subsubsection{HF08 criterion}
     

     In \cite{Happ2008}, an update criterion given by
     \begin{align}
        u_{n+1}&(D^{n}) = \argmax_{\bm{\Pi}  \in \mathcal{M}_{n+1}} \notag \\
           &\Bigl(
              \int_{\mathcal{O}} d\rho \sum_{x\in\mathcal{X}_{n+1}} p(x; \bm{\Pi}| \rho) \Delta(\hat{\rho}^{\est}_{n+1}(D^{n+1}), \rho)
           \Bigr), \label{eq:Happ}
     \end{align}
     is proposed.
     A numerical simulation is performed in \cite{Happ2008}, where the setting is that $\mathcal{O}$ is the set of 1-qubit pure states, $\mathcal{M}$ is a set of parity measurements using an ancilla system, and $\rho^{\est}$ and $\hat{\rho}^{\est}$ are maximum likelihood estimators. 
     The behavior of the average expected fidelity is numerically analyzed up to $N=20$.
     
     \subsubsection{HF11 criterion}
     
     An update criterion proposed in \cite{HappFreyberger2011} is given by
     \begin{align}
        u_{n+1}&(D^{n}) = \argmax_{\bm{\Pi}  \in \mathcal{M}_{n+1}} \notag \\
        &\Bigl( 
          - \sum_{x\in \mathcal{X}_{n+1}}\sum_{i=1}^{n} \Trace [\Pi_{i, x_{i}} \Pi_{x}] \ln \Trace [\Pi_{i, x_{i}} \Pi_{x}] 
          \Bigr),
     \end{align}
     and the estimator is defined as
     \begin{eqnarray}
        \rho^{\est}_{n}(D^n ) = \argmax_{\rho \in \mathcal{O}} \Trace [\rho \bar{\rho}(D^n ) ],\\
        \bar{\rho}(D^n ) = \frac{1}{n}\sum_{i=1}^{n}\Pi_{i,x_{i}}.
     \end{eqnarray}
     In the numerical simulations, the estimation setting is such that $\mathcal{O}$ is the set of pure states on $d$-dimensional Hilbert space $\mathcal{H}$, 
     and $\mathcal{M}_{n}$ is the set of projective measurements on $\mathcal{H}$.
     Numerical simulations of average expected fidelity are shown for $d=2,4,6,8,$ and $13$, all up to $N=50$.
         
    \subsubsection{FF00 criterion\label{subsubsec:FF00criterion}}     
    
    In \cite{FischerFreyberger2000_PLA}, an update criterion based on Bayesian estimation and Shannon entropy is proposed.
    Let $P(\rho)$ denote a prior distribution on $\mathcal{O}$.
    The update criterion is 
    \begin{align}
       u_{n+1}&(D^n ) = \argmax_{\bm{\Pi} \in \mathcal{M}_{n+1}} \notag \\
       &\Bigl(
          \sum_{x\in \mathcal{X}_{n+1}}\! \! \! \! p^{\mathrm{ave}}(x; \bm{\Pi}| D^{n}) \! \! \int_{\mathcal{O}}d\rho P(\rho | D^{n+1} ) \ln \frac{P(\rho | D^{n+1} )}{P(\rho | D^{n})}
       \Bigr) \\
       &\phantom{(D^n)}= \argmax_{\bm{\Pi} \in \mathcal{M}_{n+1}} \notag \\
       &\Bigl(
         -\int_{\rho \in \mathcal{O}}d\rho P(\rho |D^{n}) \ln P(\rho | D^{n}) + \notag \\
       &\: \sum_{x\in \mathcal{X}_{n+1}} \! \! \! \! p^{\mathrm{ave}}(x; \bm{\Pi}| D^{n})  \! \! \int_{\mathcal{O}}d\rho P(\rho | D^{n+1}) \ln P(\rho |D^{n+1})
       \Bigr) \label{eq:FF00}
    \end{align}
    where
    \begin{align}
        p^{\mathrm{ave}}(x; \bm{\Pi}| D^{n}) := \int_{\mathcal{O}} d\rho P(\rho |D^{n}) p(x; \bm{\Pi}| \rho ), \label{eq:pave}\\
        P(\rho | D^{n}) := \frac{ P(\rho)p(D^{n} | \rho) }{ \int_{\mathcal{O}} d\sigma P(\sigma )p(D^{n} | \sigma) }. \label{eq:P}
    \end{align}
    In \cite{FischerFreyberger2000_PLA}, the case in which $\mathcal{O}$ is the set of 1-qubit mixed states and $\mathcal{M}$ is the set of projective measurements is numerically analyzed up to $N=50$. The evaluation function used is the average (not expected) infidelity.  
     
    \subsubsection{HH11 criterion\label{subsubsec:HH00criterion}}     
    
     In \cite{Huszar2011},  an update criterion given by
     \begin{align}
        u_{n+1}&(D^{n}) = \argmax_{\bm{\Pi}\in \mathcal{M}_{n+1}}  
           \Bigl( 
               - \! \! \! \! \! \sum_{x \in \mathcal{X}_{n+1}} \! \! \! p^{\mathrm{ave}}(x; \bm{\Pi}| D^{n})\ln p^{\mathrm{ave}}(x; \bm{\Pi}| D^{n})  \notag \\
               &+ \int_{\mathcal{O}} d\rho P(\rho | D^{n}) \sum_{x \in \mathcal{X}_{n+1}}p(x; \bm{\Pi}| \rho)\ln p(x; \bm{\Pi}| \rho) 
           \Bigr), \label{eq:Huszar}
     \end{align}
     is proposed, where Eqs. (\ref{eq:pave}) and (\ref{eq:P}) have been used.
     From a simple calculation, we can see that the criteria defined in Eq. (\ref{eq:FF00}) and in Eq. (\ref{eq:Huszar}) are equivalent.
     This criterion involves an integration which requires high computational cost.  
     In \cite{Huszar2011}, a special technique for calculating the integral, called a sequential importance sampling method, is used in order to reduce that computational cost. 
     The authors performed numerical simulation for the case in which $\mathcal{O}$ is the set of 1-qubit mixed states and $\mathcal{M}_n$ are projective measurements up to $N=10^{4}$. 
     They also considered the case in which $\mathcal{O}$ is the set of  2-qubit states and $\mathcal{M}$ are a set of mutually unbiased bases, a set of pairwise Pauli measurements, and a set of separable measurements up to $N=10^5$.
     The evaluation function is the average expected infidelity, and it is shown that their scheme is more precise than standard quantum tomography.
      In Sec. \ref{subsubsec:AveragedMeanLosses}, we point out that our numerical results for 1-qubit show that A-optimality gives even more precise estimates than those given by Eq. (\ref{eq:Huszar}), at least from $N=100$ to $1000$.

\section{Results and analysis}\label{section:Result}

  As explained in Sec. \ref{subsec:EstimationSetting}, we consider the A-optimality criterion for 1-qubit state estimation using projective measurements.
  In Sec. \ref{subsec:AnalyticResults} we give the analytic solution, and in Sec. \ref{subsec:NumericalSimulations} we show the results of numerical simulations.

  \subsection{Analytic solution for A-optimality in 1-qubit state estimation\label{subsec:AnalyticResults}}
  
  First, we give the explicit form of the Fisher matrix for projective measurements.
  The probability distribution for the rank-1 projective measurement $\bm{\Pi}(\bm{a})$ is given by  
  \begin{eqnarray}
     p(\pm; \bm{a}| \bm{s}) &= \frac{1}{2}(1\pm \bm{s}\cdot \bm{a}),
  \end{eqnarray}
  and the Fisher matrix is 
  \begin{align}
     F(\bm{a}, \bm{s}) = &\frac{\nabla_{\bm{s}} p(+; \bm{a}|\bm{s})\nabla^T_{\bm{s}} p(+; \bm{a}|\bm{s}) }{p(+; \bm{a}|\bm{s})} \\ \nonumber
                  &+ \frac{\nabla_{\bm{s}} p(-; \bm{a}|\bm{s})\nabla^T_{\bm{s}} p(-; \bm{a}|\bm{s}) }{p(-; \bm{a}|\bm{s})} \\ 
                = &\frac{\bm{a}\bm{a}^{T}}{1-(\bm{a}\cdot \bm{s})^2}.
  \end{align}
 In this case, Eq. (\ref{eq:DefAopt}) is rewritten in the Bloch vector representation as 
 \begin{align}
   \bm{a}^{\Aopt}_{n+1} := \argmin_{\bm{a}\in\mathcal{A}} \trace \big[ H(\hat{\bm{s}}^{\est}_{n}) \{ \tilde{F}_{n}(\bm{a}^{n}, \hat{\bm{s}}^{\est}_{n} |D^{n}) + F(\bm{a}, \hat{\bm{s}}^{\est}_{n}) \}^{-1}\big]. \label{eq:Aopt1qubit}
 \end{align}
 We present the analytic solution of Eq.(\ref{eq:Aopt1qubit}) in the form of the following theorem. 
  \begin{theorem}\label{theorem:1}
    Given  a sequence of data $D^{n} = \{ (\bm{a}_{1}, x_{1}), \ldots, (\bm{a}_{n}, x_{n})\}$, the $n$-th estimate $\hat{\bm{s}}^{\est}_{n}$, and a real positive matrix $H$, the A-optimal POVM Bloch vector is given by
\begin{align}
    \bm{a}^{\Aopt}_{n+1} = \frac{B_{n}\bm{e}_{\mathrm{min}}(C_{n})}{\| B_{n}\bm{e}_{\mathrm{min}}(C_{n})\|},\label{eq:AoptTheorem1-1}
\end{align}
where
\begin{align}
   B_{n} &= \sqrt{ \tilde{F}_{n}( \bm{a}^{n}, \hat{\bm{s}}^{\est}_{n} | D^{n}) H( \hat{\bm{s}}^{\est}_{n})^{-1} \tilde{F}_{n}(\bm{a}^{n}, \hat{\bm{s}}^{\est}_{n}|D^{n} )}, \label{eq:AoptTheorem1-2}\\
   C_{n} &= B_n (I - \hat{\bm{s}}^{\est}_{n} \hat{\bm{s}}^{\est T}_{n}+\tilde{F}_{n}(\bm{a}^{n}, \hat{\bm{s}}^{\est}_{n} |D^{n})^{-1}) B_n, \label{eq:AoptTheorem1-3}
\end{align}
$\bm{e}_{\mathrm{min}}(C_{n})$ is the eigenvector of the matrix $C_{n}$ corresponding to the minimal eigenvalue, and $I$ is the identity in the parameter space.
\end{theorem}
We give the proof of Theorem \ref{theorem:1} in Appendix \ref{sec:appendixA}. 

In Eq. (\ref{eq:AoptTheorem1-3}), the inverse of the matrix $\tilde{F}_{n}$ appears.
In the proof of Theorem \ref{theorem:1}, the invertibility of $\tilde{F}_{n}$ is assumed.
The invertibility of $\tilde{F}_{N}$ is equivalent to the condition that $\bm{a}^{N}$ is a basis of $\mathbb{R}^{3}$.
When we choose the second and third measurements, $\tilde{F}_{1}$ and $\tilde{F}_{2}$ are not invertible.
Thus the update scheme does not apply to these steps, and the choices are arbitrary.
One simple choice is to perform $\sigma_{1}$-, $\sigma_{2}$-, and $\sigma_{3}$-projective measurements at the first, second and third trials respectively, and this can be shown to satisfy Theorem \ref{theorem:1} as follows.
The choice of the first measurement is always arbitrary, and we choose $\bm{a}_{1}=(1,0,0)^{T}$, a $\sigma_{1}$-projective measurement.
Then for any true Bloch vector $\bm{s}$ the rank of $\tilde{F}_{1}$ is $1$, and if we interpret the inverse matrix in Eq. (\ref{eq:AoptTheorem1-3}) as a generalized inverse matrix, $C_{1}$ is a rank $1$ matrix with minimal eigenvalues $0$.
The supports of $\tilde{F}_{1}$, $B_{1}$, and $C_{1}$ are the span of $\{ \bm{a}_{1}\}$.
Therefore $B_{1}\bm{e}_{\mathrm{min}}(C_{1})$ is an arbitrary vector in the $2$-dimensional space spanned by $(0,1,0)^{T}$ and $(0,0,1)^T$,
and we choose $\bm{a}_{2}=(0,1,0)^{T}$.
Then using the same logic, the third measurement is fixed to $\bm{a}_{3}=(0,0,1)^T$.

From the explicit formulae of the squared Hilbert-Schmidt distance and infidelity in Eqs. (\ref{eq:HSdistance2_para}) and (\ref{eq:Infidelity_para}), we have 
\begin{align}
   \Delta^{\mathrm{HS}} (\bm{s} , \bm{s}^{\prime} ) &= (\bm{s}^{\prime} - \bm{s} )^{T}\frac{1}{4}I(\bm{s}^{\prime} - \bm{s} ), \\
   \Delta^{\mathrm{IF}} (\bm{s} , \bm{s}^{\prime} ) &= (\bm{s}^{\prime} - \bm{s} )^{T}\frac{1}{4}\Bigl(I + \frac{\bm{s}\bm{s}^{T}}{1-\| \bm{s}\|^2} \Bigr)(\bm{s}^{\prime} - \bm{s} ) \nonumber \\ 
   &\quad + O(\|\bm{s}^{\prime} - \bm{s} \|^3). 
\end{align}
Therefore when we use the Hilbert-Schmidt distance as our loss function, we substitute $H^{\mathrm{HS}}(\bm{s}) := \frac{1}{4}I$ and $H^{\mathrm{HS}}(\bm{s})^{-1} = 4I$ into Eqs.(\ref{eq:AoptTheorem1-1}), (\ref{eq:AoptTheorem1-2}), and (\ref{eq:AoptTheorem1-3}) to obtain
\begin{align}
   B_{n} &= \tilde{F}_{n}(\bm{a}^{n}, \hat{\bm{s}}^{\est}_{n}|D^{n}), \\
   C_{n} &= \tilde{F}_{n}(\bm{a}^{n}, \hat{\bm{s}}^{\est}_{n}|D^{n}) (I - \hat{\bm{s}}^{\est}_{n} \hat{\bm{s}}^{\est T}_{n}) \tilde{F}_{n}(\bm{a}^{n}, \hat{\bm{s}}^{\est}_{n}|D^{n}) \nonumber \\
            &\quad+  \tilde{F}_{n}(\bm{a}^{n}, \hat{\bm{s}}^{\est}_{n}|D^{n}), 
\end{align}
and we do not need to explicitly calculate the inverse or square root matrices for A-optimality.
On the other hand, when our loss function is the infidelity, we must use $H^{\mathrm{IF}}(\bm{s}) := \frac{1}{4}\Bigl(I + \frac{\bm{s}\bm{s}^{T}}{1-\| \bm{s}\|^2} \Bigr)$ and $H^{\mathrm{IF}}(\bm{s})^{-1} = 4(I - \bm{s}\bm{s}^{T})$.

  \subsection{Numerical simulation \label{subsec:NumericalSimulations}}
  
    We performed Monte Carlo simulations of the following four experimental designs described in detail below;
    A-optimal adaptive scheme for the squared Hilbert-Schmidt distance, the same for infidelity, XYZ repetition, and uniformly random selection.
      
    A-optimality for the squared Hilbert-Schmidt distance is the adaptive scheme defined by Eq.(\ref{eq:Aopt1qubit}) with $H=H^{\mathrm{HS}}$.
    Similarly, A-optimality for the infidelity is that with $H=H^{\mathrm{IF}}$.
    As explained in the previous subsection, the choice of measurement Bloch vectors at the first and second trials is arbitrary; we choose $\bm{a}_{1}=(1,0,0)^{T}$ and $\bm{a}_{2}=(0,1,0)^T$, i.e., at the first trial we perform the projective measurement of $\sigma_{1}$, and that of $\sigma_{2}$ at the second --- the third trial is automatically the projective measurement of $\sigma_{3}$, corresponding to $\bm{a}_{3}=(0,0,1)^T$.
    The XYZ repetition scheme is nonadaptive, in which we repeat the measurements of $\sigma_1$, $\sigma_2$, and $\sigma_3$, corresponding to standard quantum state tomography.
    Uniformly random selection is also nonadaptive, where at each trial we choose the next measurement direction randomly on the Bloch surface, according to the SO(3) Haar measure.
    For consistency with the other three schemes, we fix the first, second and third measurements to be the projective measurements of $\sigma_1 , \sigma_2 , \sigma_3$, respectively, and randomly select directions from the fourth trial on.
    
    We choose a maximum likelihood estimator in all four experimental designs.
    It is known that the estimators minimizing $\bar{\Delta}^{\mathrm{HS} \mathrm{ave}}$ and $\bar{\Delta}^{\mathrm{IF} \mathrm{ave}}$ are Bayesian estimators \cite{Bagan2004, RobinNJP2010}, but the integrations necessary for Bayesian estimation take too much computation time. 
    For the two A-optimality criteria, we choose both the real and the dummy estimators to be maximum likelihood, $\bm{s}^{\mathrm{est}} = \hat{\bm{s}}^{\mathrm{est}} = \bm{s}^{\mathrm{ML}}$.
    We used a Newton-Raphson method to solve the (log-)likelihood equation and the completely mixed state $\bm{s}=\bm{0}$ as the initial point of the iterative method.
    When a search point came out of the Bloch sphere during the procedure, we chose the previous point (included in the sphere) as the estimate. 
        
    In the following subsections, we show the plots for two loss functions; the squared Hilbert-Schmidt distance $\Delta^{\mathrm{HS}}$ and infidelity $\Delta^{\mathrm{IF}}$.
    The average expected losses $\bar{\Delta}^{\mathrm{ave}}_{N}$ are shown in Sec. \ref{subsubsec:AveragedMeanLosses}, and pointwise expected losses $\bar{\Delta}_{N}$ are shown in Sec. \ref{susubsec:PointwiseMeanLosses}.
    In the both subsections, the line styles are fixed as follows:
      solid (black) line for A-optimality for the squared Hilbert-Schmidt distance (AHS),
      dashed (red) line for A-optimality for the infidelity (AIF),
      chain (blue) line for XYZ repetition (XYZ),
      Dotted (green) line for Uniformly random selection (URS).

  \subsubsection{Average expected losses\label{subsubsec:AveragedMeanLosses}}
   
   \begin{figure*}[tbhp]
       \includegraphics[width =0.95\linewidth]{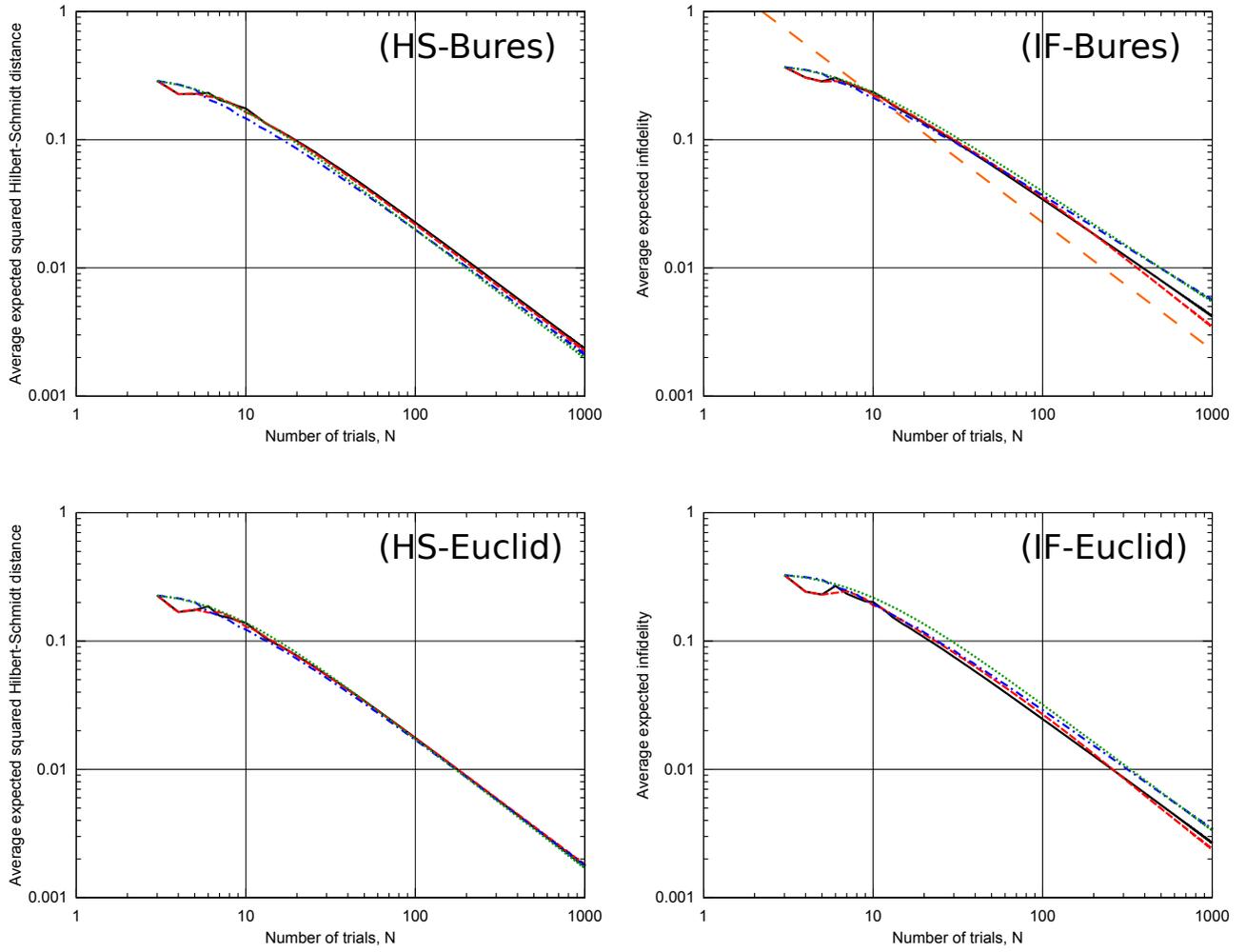}%
       \caption{\label{fig:AveragedExpectedLosses} Average expected loss $\bar{\Delta}^{\mathrm{ave}}_{N}(u^{N}, \bm{s}^{\mathrm{est}})$ against the number of measurement trials $N$: 
  (HS-Bures) $\bar{\Delta}^{\mathrm{HS} \mathrm{ave}}_{N}$ integrated via the Bures distribution $\mu_{\mathrm{Bures}}$,
  (HS-Euclid) $\bar{\Delta}^{\mathrm{HS} \mathrm{ave}}_{N}$ integrated via the Euclidean distribution $\mu_{\mathrm{Euclid}}(\bm{s}) = 3/4\pi$,    
  (IF-Bures) $\bar{\Delta}^{\mathrm{IF} \mathrm{ave}}_{N}$ integrated via $\mu_{\mathrm{Bures}}$,
  and (IF-Euclid) $\bar{\Delta}^{\mathrm{IF} \mathrm{ave}}_{N}$ integrated via $\mu_{\mathrm{Euclid}}$.
  The dashed spaced (orange) line in (IF-Bures) is the bound of separable (including adaptive) schemes derived in \cite{Bagan2006PRL}.
  The number of measurement trials $N_{\mathrm{max}}$ is $1000$, the number of sequences used for the calculation of the statistical expectation values $N_{\mathrm{mean}}$ is $1000$, and the number of sample points used for the Monte Carlo integration $N_{\mathrm{MC}}$ is $3200$.}
   \end{figure*}   

  We analyse the average behaviour of the estimation errors over the Bloch sphere.
  The integration for averaging is approximated by a Monte Carlo routine, and 
  the statistical expectation is approximated by an arithmetric mean using pseudo-random numbers.

  Figure \ref{fig:AveragedExpectedLosses} shows the average expected loss functions $\bar{\Delta}^{\mathrm{ave}}_{N}$ against the number of trials $N$ (the horizontal and vertical axes are both logarithmic scale):
  (HS-Bures) $\bar{\Delta}^{\mathrm{HS} \mathrm{ave}}_{N}$ integrated via the Bures distribution $\mu_{\mathrm{Bures}}$,
  (HS-Euclid) $\bar{\Delta}^{\mathrm{HS} \mathrm{ave}}_{N}$ integrated via the Euclidean distribution $\mu_{\mathrm{Euclid}}(\bm{s}) = 3/4\pi$,  
  (IF-Bures) $\bar{\Delta}^{\mathrm{IF} \mathrm{ave}}_{N}$ integrated via $\mu_{\mathrm{Bures}}$,
  and (IF-Euclid) $\bar{\Delta}^{\mathrm{IF} \mathrm{ave}}_{N}$ integrated via $\mu_{\mathrm{Euclid}}$.
  Fig. \ref{fig:AveragedExpectedLosses} (HS-Bures) and (HS-Euclid) shows that the estimation errors of the four experimental designs are almost equivalent from the viewpoint of the squared Hilbert-Schmidt distance.
  As depicted in (HS-Bures), the estimation errors of the two A-optimality schemes are slightly larger than the other nonadaptive schemes; as we show in the next subsection (pointwise analysis), this gap decreases as $N$ becomes larger.
  On the other hand, Fig. \ref{fig:AveragedExpectedLosses} (IF-Bures) and (IF-Euclid) show the explicit gap between the adaptive and nonadaptive schemes.
  The gradients of the curves begin to differentiate from around $N=100$, and as depicted in (IF-Bures), the gradients of XYZ and URS are almost $-3/4$ around $N=1000$.
  This means that the average expected infidelity behaves as $O(N^{-3/4})$ and is consistent with the result of the asymptotic analysis presented in \cite{Bagan2004}.
  On the other hand, the gradients of AHS and AIF are greater than the nonadaptive limit $-3/4$, indicating that AHS and AIF make good use of adaptive resources.
  Around $N=1000$ the gradient of AIF is almost $-1$, which is the bound for adaptive experimental designs \cite{Bagan2006PRL}. 

  Let us compare the estimation errors of A-optimality and the HH11 criteria explained in Sec. \ref{subsubsec:FF00criterion}.
  From Fig. \ref{fig:AveragedExpectedLosses} (IF-Bures), the average expected infidelity of AHS and AIF are $4.2\times 10^{-3}$ and $3.5\times 10^{-3}$ at $N=1000$.
   On the other hand, the corresponding amount for the HH11 criterion can be estimated roughly from Fig. 2 (a) in \cite{Huszar2011} to be $7.0\times 10^{-3}$.
   This implies that for 1-qubit state estimation, the average expected infidelity of the A-optimality criterion is about two-times smaller than that of Eq. (\ref{eq:Huszar}), at least around $N=1000$.

   \begin{figure*}[tbhp]
       \includegraphics[width =0.85\linewidth]{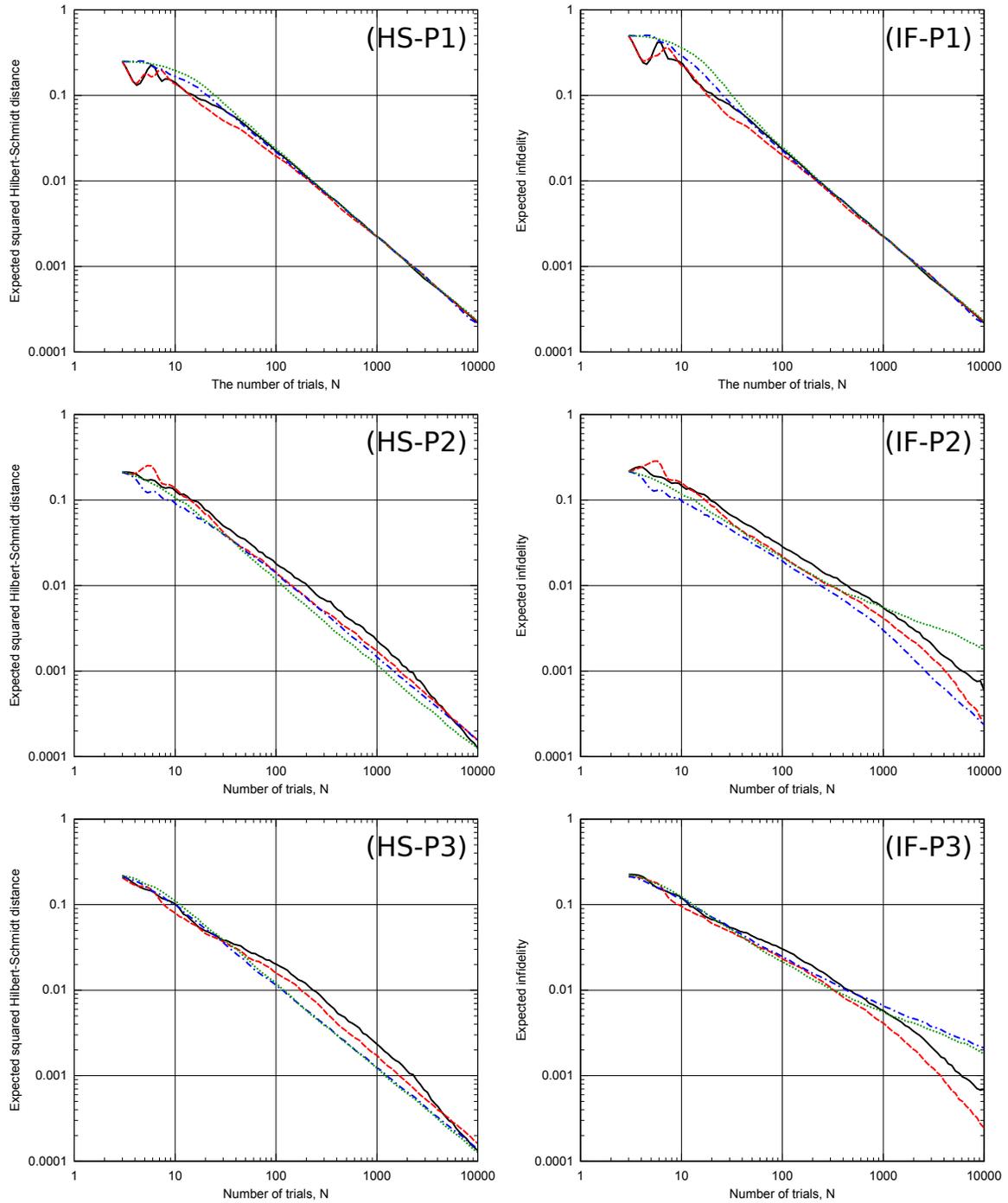}%
       \caption{\label{fig:PointwiseMeanLosses} Pointwise expected loss $\bar{\Delta}_{N}(u^{N}, \bm{s}^{\mathrm{est}} |\bm{s})$ against the number of trials $N$ (the horizontal and vertical axes are both logarithmic scale):  (HS-P1), (HS-P2), and (HS-P3) are the expected squared Hilbert-Schmidt distances for $\bm{s}$ given by $(r, \theta , \phi) = (0,0,0),\ (0.99,0,0),\ (0.99, \pi/4 , \pi/4)$, and (IF-P1), (IF-P2), and (IF-P3) are the expected infidelities for the same three true states, respectively. The number of measurement trials $N_{\mathrm{max}}$ is $10000$, and the number of sequences used for the calculation of statistical expectation values $N_{\mathrm{mean}}$ is $1000$.}
   \end{figure*} 

  \subsubsection{Pointwise expected losses\label{susubsec:PointwiseMeanLosses}}

   Next, we analyse the behaviour of the estimation errors at several true Bloch vectors $\bm{s}$. 
   Figure \ref{fig:PointwiseMeanLosses} shows the pointwise expected loss functions $\bar{\Delta}_{N}(u^{N} |\bm{s})$ against the number of trials $N$ (the horizontal and vertical axes are both logarithmic scale):  
     (HS-P1), (HS-P2), and (HS-P3) are plots of the expected squared Hilbert-Schmidt distances for $\bm{s}$ given by $(r, \theta , \phi) = (0,0,0),\ (0.99,0,0),\ (0.99, \pi/4 , \pi/4)$, and
     (IF-P1), (IF-P2), and (IF-P3) are the expected infidelities for the same three true states, respectively.
     
   As depicted in (HS-P1) and (IF-P1), the estimation errors of all four schemes are almost equivalent for the completely mixed state, $\bm{s}=\bm{0}$.      
   As the Bloch radius $r$ becomes larger, the differences between the four schemes become clearer.
   Figure \ref{fig:PointwiseMeanLosses} (HS-P2) and (HS-P3) are the plots of the expected squared Hilbert-Schmidt distances at a high purity point, $r=0.99$.
   In the region of $N=10$ to around $7000$, the squared Hilbert-Schmidt error of the two adaptive schemes is larger than that of the two nonadaptive schemes.
   In particular, the error of AHS is larger that that of AIF; this might seem strange, but in the region of $N\ge 7000$, the error of AHS becomes smaller than that of AIF, indeed it eventually becomes the smallest of the four schemes.
   We believe that there are two reasons for A-optimality's large error for small $N$.   
   First, the A-optimality criterion is based on an asymptotic theory of statistical estimation.
   When the number of measurement trials $N$ is small, the Cram\'{e}r-Rao bound is not necessary suitable for characterizing estimation errors.
   Second, it uses a dummy estimator in the measurement update.
   When $N$ is small, $\hat{\bm{s}}^{\mathrm{est}}_{N}$ is not a good estimate, and thus the choice of the next measurements can be unreliable.
   Of course, when $N$ becomes sufficiently large, both of these problems are alleviated.
    
   The gap between the estimation errors of adaptive and nonadaptive schemes becomes smaller as $N$ becomes larger in (HS-P2) and (HS-P3), while it grows in (IF-P2) and (IF-P3).
    Only the XYZ scheme changes dramatically between (IF-P2) and (IF-P3); the other three schemes do not because AHS, AIF, and URS are invariant under rotation of the true Bloch vector (for very small $N$, there are  differences, and these are because the first three measurements are fixed to $\sigma_{1}, \sigma_{2}, \sigma_{3}$-projective measurements and not rotationally invariant).
   Figure \ref{fig:PointwiseMeanLosses} (IF-P2) is the case in which the directions of the measurement and the true Bloch vector are matched (to $(0,0,1)$).
   In this case, XYZ is the best scheme, exhibiting the smallest estimation error. 
   Around $N=10000$, the estimation error of AIF becomes as small as that of XYZ.
   That of AHS is smaller than URS, but larger than the other two schemes.
   We believe that this is because the selected Hessian matrix $H^{\mathrm{HS}}$ used in the update routine is unsuitable for the loss function $\Delta^{\mathrm{IF}}$ in (IF-P2) (and (IF-P3)).
   Figure \ref{fig:PointwiseMeanLosses} (IF-P3) is the case in which the directions of the measurement and the true Bloch vector are the most discrepant (for a fixed purity).   
   In this case, the estimation errors of XYZ and URS are almost the same and behave as $O(N^{-1/2})$, and those of the adaptive schemes are smaller than those of the nonadaptive ones, (this behavior of expected infidelity for i.i.d. measurements is discussed in \cite{Bagan2006PRL, Burgh2008}, and a detailed analysis will appear in \cite{Sugiyama2012}).
   When we consider the whole Bloch sphere, of course the cases in which the direction of XYZ measurements and the Bloch vector are matched are few, and therefore the average expected infidelities of AHS and AIF are smaller than those of XYZ and URS.
   This also indicates that the adaptive schemes have better worst-case performance (lower $\bar{\Delta}^{\mathrm{max}}_{N}$, Eq. (\ref{eq:maximalloss})) than the nonadaptive schemes.

  \subsubsection{Purity dependence}
  
   \begin{figure}[htbp]
       \includegraphics[width =0.85\linewidth]{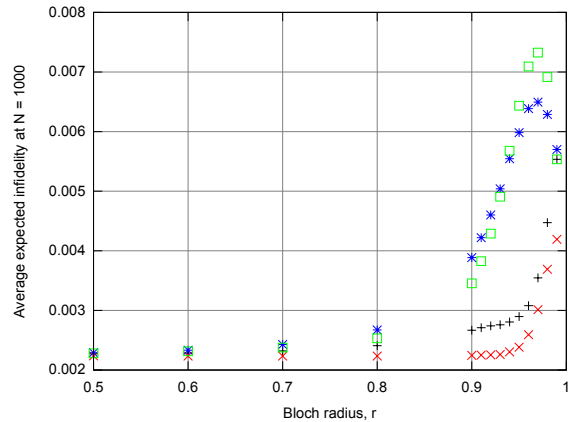}%
       \caption{\label{fig:Results_purity} Purity dependence of average expected infidelity at $N=1000$. 
       Cross (black) for AHS,  saltire (red) for AIF, asterisk (blue) for XYZ, and square (green) for URS.
       The number of sequences used for the calculation of the statistical expectation values $N_{\mathrm{mean}}$ is $1000$, and the number of sample points used for the Monte Carlo integration $N_{\mathrm{MC}}$ is $500$ for each Bloch radius $r$.}
   \end{figure}   
   
   Figure \ref{fig:Results_purity} shows the purity dependence of the average expected infidelity at $N=1000$.
   The average is taken over all directions $\theta$ and $\phi$ for each Bloch radius $r$.
   It indicates that the average expected infidelities of the two adaptive schemes are smaller than those of the two nonadaptive schemes.
   The appearance of peaks for XYZ and URS is discussed in Appendix~\ref{sec:PurityXYZURS}.   

  \subsubsection{Measurement sequences}
  
   \begin{figure*}[htbp]
       \includegraphics[width =0.85\linewidth]{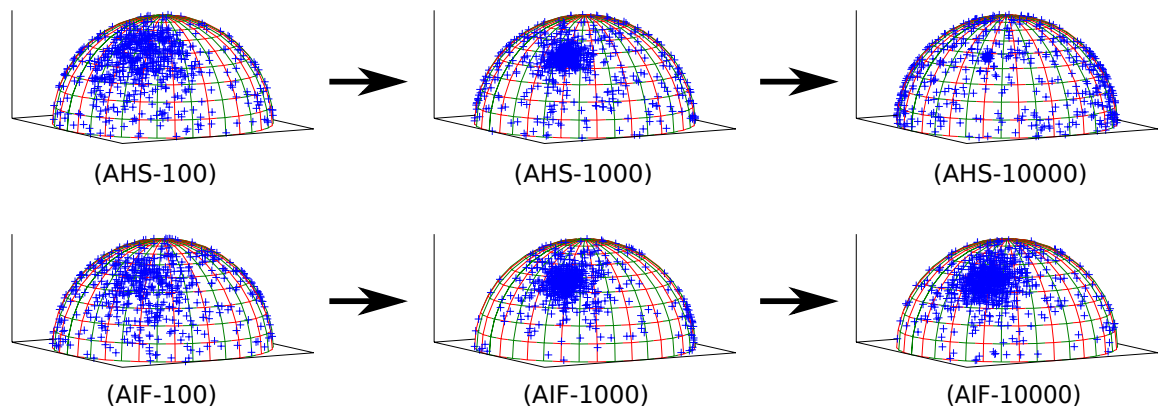}%
       \caption{\label{fig:MeasurementAnalysis} Distribution of measurement Bloch vectors at $N=100$ (left column), $1000$ (middle column), and $10000$ (right column) for $900$ runs; The true state is $(r, \theta , \phi) = (0.99, \pi/4 , \pi/4)$. The upper three plots (AHS-100), (AHS-1000), and (AHS-10000) are AHS while the lower three (AIF-100), (AIF-1000), and (AIF-10000) are AIF.}
   \end{figure*}   
   
   Figure \ref{fig:MeasurementAnalysis} is a plot of the measurement Bloch vectors at $N=100$ (left column), $1000$ (middle column), and $10000$ (right column) for $900$ runs.
   The true state is $(r, \theta , \phi) = (0.99, \pi/4 , \pi/4)$, and the upper three subplots are AHS while the lower three are AIF.
   Figure \ref{fig:MeasurementAnalysis} shows that the measurement Bloch vectors are clustered around the true state, with some interesting behaviour at $N=10000$.
   In (AHS-10000), the measurement directions are clustered very narrowly at the true state and also around the great circle that it defines.
   In (AIF-10000), on the other hand, the directions are clustered widely around the true state.
   This is due to the difference between the loss functions employed in the update routine, namely squared Hilbert-Schmidt distance in the former and infidelity in the latter.  
    We mention that for a completely mixed true state, the measurement Bloch vectors are distributed randomly on the Bloch sphere for large $N$.

\section{Discussion}\label{section:discussion}

  \subsection{Implementation}

   \begin{figure}[tbhp]
       \includegraphics[width =0.95\linewidth]{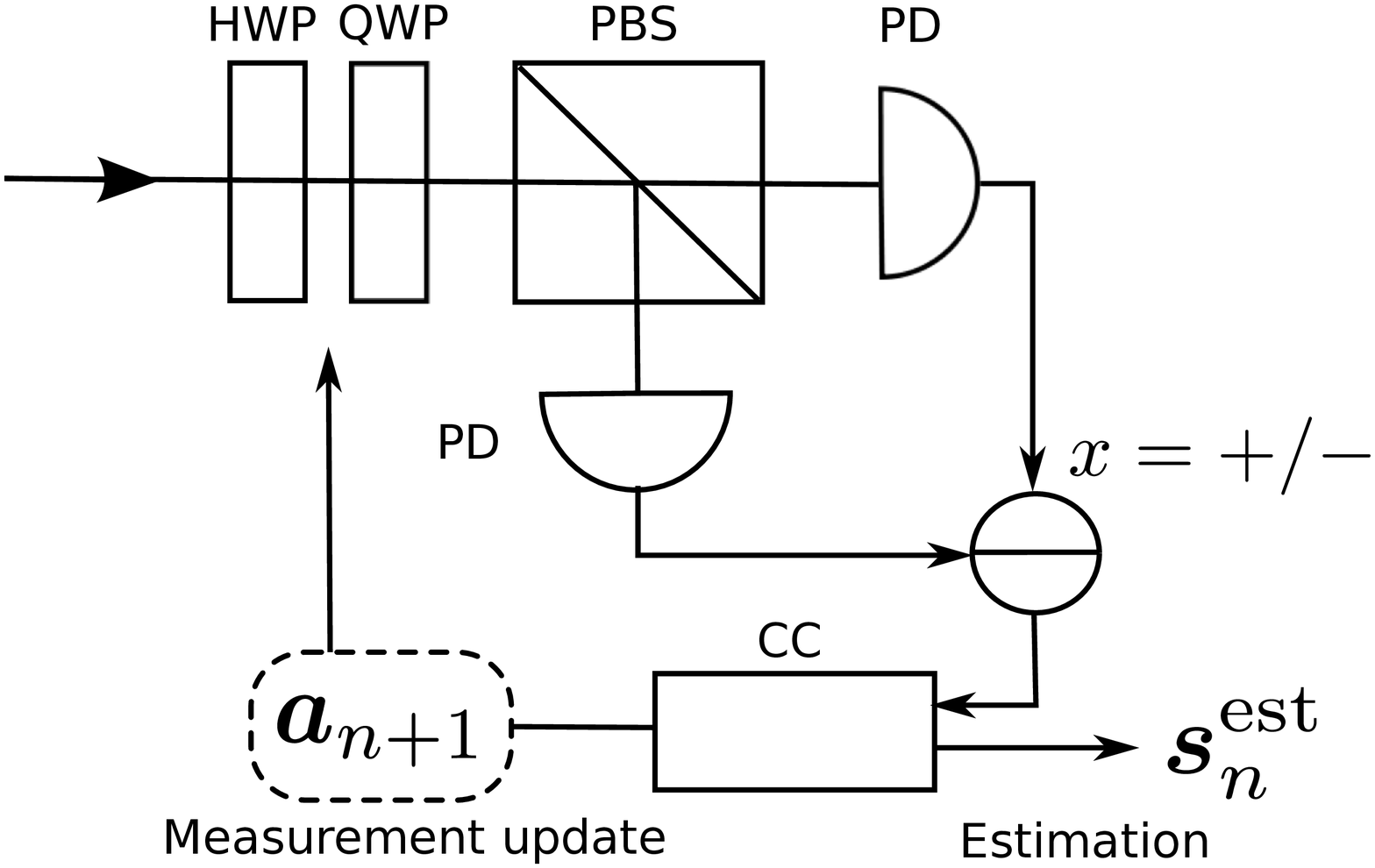}%
       \caption{\label{fig:PolarizationImplimentation} An implementation of adaptive projective measurements for single photon polarization qubits in quantum optics. HWP and QWP are half and quarter wave plates, PBS is a polarization beam splitter, PD are photodetectors, and CC denotes classical computation. The direction of the projective measurement are adapted by changing the waveplate angles.}
   \end{figure} 
    
    There are two main issues when considering the practical implementation of an adaptive scheme, namely the ease with which measurement updates can be made in the apparatus, and the time required to compute those updates.  In quantum optics, projective measurements and single qubit rotations are standard tools in quantum information processing experiments.
     Figure \ref{fig:PolarizationImplimentation} illustrates a simple implementation example for a one photon polarization system.  In this regard, the first issue is not a problem  --- in general, of course it will depend on the experimental state of the art.

       
  %
  \subsection{Generalization to higher dimensional systems}
  
  In order to compare the performance of the A-optimality criterion to the other update schemes, we have considered 1-qubit states as the estimation objective.
  Current and future quantum information processing is concerned with higher dimensional estimation objectives, not only states but also processes.
  In 1-qubit state estimation, we can reduce the computational cost for A-optimality by using the analytic solution of Theorem \ref{theorem:1}, but as we see in Appendix \ref{sec:appendixA}, the techniques used to derive that solution depend on the properties of 1-qubit states and projective measurements. 
  A-optimality in higher dimensional systems will need a new solution, or must deal with the increasing complexity of the nonlinear minimization problem.
  One possible approach is to place constraints on the measurement class $\mathcal{M}_{n}$.
  Instead of considering a continuous set of measurement candidates, we could consider a discrete set. 
  One expects that the resulting discrete minimization problem would be much simpler.
  If the number of discrete measurement candidates is too small however, the estimation error could be worse than standard quantum tomography.
  The relation between the reduction in computational cost and the (probable) increase in estimation error by introducing such discrete minimization is an open problem.

\section{Summary}\label{section:summary}

In this paper, we considered adaptive experimental design and applied a measurement update method known in statistics as the A-optimality criterion to 1-qubit mixed state estimation using arbitrary rank-1 projective measurements. 
We derived an analytic solution of the A-optimality update procedure in this case, reducing the complexity of measurement updates considerably.
Our analytic solution is applicable to any case in which the loss function can be approximated by a quadratic function to least order.
We performed Monte Carlo simulation of this and several nonadaptive schemes in order to compare the behaviour of estimation errors for a finite number of measurement trials.
We compared the average and pointwise expected squared Hilbert-Schmidt distance and infidelity of the following four measurement update criteria: 
   A-optimality for the squared Hilbert-Schmidt distance (AHS), 
   A-optimality for the infidelity (AIF), 
   repetition of three orthogonal projective measurements (XYZ), and 
   uniformly random selection of projective measurements (URS).
The numerical results showed that AHS and AIF give more precise estimates than URS and XYZ which corresponds to standard quantum tomography with respect to expected infidelity.

\begin{acknowledgments}
   T.S. would like to thank Fuyuhiko Tanaka for helpful discussion on mathematical statistics and Terumasa Tadano for useful advice on numerical simulation.  
   This work was supported by JSPS Research Fellowships for Young Scientists (22-7564) 
   and Project for Developing Innovation Systems of the Ministry of Education, Culture, Sports, Science and Technology (MEXT), Japan.
\end{acknowledgments}

\appendix
\section{Proof of Theorem \ref{theorem:1}}\label{sec:appendixA}

We give the proof of Theorem \ref{theorem:1}.
First, we introduce a lemma about matrix inverses.
\begin{lemma}\cite{MatrixAnalysisText1}
 Let $V$ denote a $k\times k$ invertible matrix.
 Let us consider a matrix $W = V +\bm{v}\bm{v}^{T}$,
 where $\bm{v}$ is a $k$-dimensional vector.
 If $W$ is not singular, then
 \begin{align}
    W^{-1} = V^{-1} -\frac{V^{-1}\bm{v}\bm{v}^{T}V^{-1}}{1+\bm{v}^{T}V^{-1}\bm{v}}.\label{eq:Inverse2}
 \end{align} 
\end{lemma}

 By substituting $\bm{v}=\bm{a} / \sqrt{1-(\bm{a}\cdot\bm{s})}$ into Eq.(\ref{eq:Inverse2}) (in our case $k=3$ and $V = \tilde{F}_{n}$), we obtain
  \begin{align}
     \{ V+F(\bm{a},\bm{s}) \}^{-1} 
      &= V^{-1}- \frac{V^{-1} \bm{a}\bm{a}^{T} V^{-1}}{1-(\bm{a}\cdot \bm{s})^2 + \bm{a}^{T}V^{-1}\bm{a}},
  \end{align}
  and 
  \begin{align}
     \mbox{tr}&[H(\bm{s}) \{ V + F(\bm{a},\bm{s}) \}^{-1} ] =  \notag \\
     &\mbox{Tr}[H(\bm{s}) V^{-1}] - \frac{\bm{a}^{T} V^{-1}H(\bm{s})V^{-1} \bm{a}}{1-(\bm{a}\cdot \bm{s})^2 + \bm{a}^{T} V^{-1}\bm{a}}.\label{eq:A3}
  \end{align}
  The first term of the RHS in Eq.(\ref{eq:A3}) is independent of  $\bm{a}$ and therefore we obtain 
  \begin{align}
     \argmin_{\bm{a}\in \mathcal{A}}\mbox{tr}[H(\bm{s}) \{ V + F(\bm{a}, \bm{s}) \}^{-1} ]\notag \\
     = \argmax_{\bm{a}\in \mathcal{A}}\frac{\bm{a}^{T} V^{-1}H(\bm{s}) V^{-1}\bm{a}}{\bm{a}^{T}(I-\bm{s}\bm{s}^{T} + V^{-1})\bm{a}} \\
     = \argmin_{\bm{a}\in \mathcal{A}} \frac{\bm{a}^{T}(I-\bm{s}\bm{s}^{T} + V^{-1})\bm{a}}{\bm{a}^{T} V^{-1}H(\bm{s}) V^{-1}\bm{a}},
  \end{align}
  where we used the relation $1 = \bm{a}^{T}I \bm{a}$.
  Let us introduce a vector
  \begin{eqnarray}
     \bm{b} := \frac{ \sqrt{V^{-1} H(\bm{s}) V^{-1}} \bm{a} }{ \| \sqrt{V^{-1} H(\bm{s}) V^{-1}} \bm{a} \| }.\label{eq:A6}
  \end{eqnarray}
  Note that $\bm{b}$ and $\bm{a}$ take values in the same set, so that the vector $\bm{a}$ can be represented in terms of $\bm{b}$ as
  \begin{eqnarray}
     \bm{a} = \frac{ \sqrt{V H(\bm{s})^{-1} V} \bm{b} }{ \| \sqrt{V H(\bm{s})^{-1} V} \bm{b} \| }.\label{eq:A7}
  \end{eqnarray}  
  Then the minimization function is represented by using $\bm{b}$ as
  \begin{align}
     &\frac{\bm{a}^{T}(I-\bm{s}\bm{s}^{T} + V^{-1})\bm{a}}{\bm{a}^{T} V^{-1}H(\bm{s}) V^{-1}\bm{a}}  = \notag \\
     & \bm{b}^{T}\sqrt{VH(\bm{s})^{-1}V}(I-\bm{s}\bm{s}^{T} + V^{-1})\sqrt{VH(\bm{s})^{-1}V}\bm{b}.\label{eq:A8}
   \end{align}
   The vector $\bm{b}$ minimizing Eq.(\ref{eq:A8}) is the eigenvector with the minimal eigenvalue of the matirx 
   \begin{align}
      C:=\sqrt{VH(\bm{s})^{-1}V}(I-\bm{s}\bm{s}^{T} + V^{-1})\sqrt{VH(\bm{s})^{-1}V},\label{eq:A9}
   \end{align}
   i.e., $\bm{b}=\bm{e}_{\mathrm{min}}(C)$.
   By substituting $V=\tilde{F}_{n}$ and $\bm{s}=\hat{\bm{s}}^{\mathrm{est}}_{n}$ into Eqs. (\ref{eq:A7}) and (\ref{eq:A9}), we obtain Theorem \ref{theorem:1}.
  $\blacksquare$

  \section{Proof of a generalized Cram\'{e}r-Rao inequality \label{sec:ProofCRineq}}
  
  We give a proof of Eq.(\ref{eq:gCRineq}). 
  We consider a general probability distribution $\{ p(y|\theta ) \}_{y\in\mathcal{Y}}$, where $\mathcal{Y}$ is a set of outcomes and $\theta \in \Theta \subset \mathbb{R}^k$.
  It does not necessarily obey Born's rule. 
  We assume differentiability of a sufficient order with respect to $\theta$.
  From the definition of probability distributions, we obtain
  \begin{eqnarray}
     1&=& \sum_{y \in \mathcal{Y}} p(y |\theta), \\
     0&=& \sum_{y\in \mathcal{Y}} \nabla_{\theta}p(y|\theta) = \sum_{y\in \mathcal{Y}} p(y|\theta)\nabla_{\theta}\ln p(y|\theta),\label{eq:B2}
  \end{eqnarray}
  where we assumed that $\forall \theta\ \mbox{and}\ \forall y \in \mathcal{Y} , p(y |\theta )>0$. 
  This assumption is valid for all full rank density matrices in any finite dimensional system.
  The contrapositive is that there can exist non full rank density matrices which do not satisfy the assumption.
  This is the reason why we restrict our estimation objective to mixed states, the interior of the Bloch sphere in Sec. \ref{subsec:EstimationSetting}.
  Let us define a $k\times k$ matrix $G$ as
  \begin{eqnarray}
     G :&=& \nabla_{\theta} \sum_{y\in\mathcal{Y}} p(y|\theta) \theta^{\est}(y)^T \\
          &=& \sum_{y\in\mathcal{Y}} p(y|\theta) \nabla_{\theta} \ln p(y|\theta) (\theta^{\est}(y) - \theta)^T ,
  \end{eqnarray}
  where we used Eq.(\ref{eq:B2}).
  For any vectors $u$ and $w$ in $\mathbb{R}^{k}$, we obtain
  \begin{eqnarray}
     &&( u^T G w)^2 \notag \\
     &=& \Bigl( \sum_{y\in \mathcal{Y}}p(x|\theta)[u^T \nabla_{\theta}\ln p(x|\theta)] [(\theta^{\est}(y)-\theta)^{T}w] \Bigr)^2 \\
     &\le&\Bigl(\sum_{y\in\mathcal{Y}}p(y|\theta)[u^T \nabla_{\theta}\ln p(y|\theta)]^2 \Bigr) \notag \\
      & & \ \ \times  \Bigl( \sum_{y^{\prime}\in\mathcal{Y}} p(y^{\prime}|\theta)[(\theta^{\est}(y)-\theta)^{T}w]^2 \Bigr)  \notag \\
     &=& (u^T F u) (w^T E w),
   \end{eqnarray}
   where
   \begin{eqnarray}
      E:= \sum_{y\in\mathcal{Y}} p(y|\theta)(\theta^{\est}(y)-\theta)(\theta^{\est}(y)-\theta)^T ,\\
      F:= \sum_{y\in\mathcal{Y}}p(y|\theta) \nabla_{\theta}\ln p(y|\theta)  \nabla_{\theta}^T \ln p(y|\theta).
   \end{eqnarray}
   Therefore $\forall u, w$, we obtain 
   \begin{eqnarray}
      w^T E w \ge \frac{  u^T G w w^T G^{T} u} {u^T F u}. \label{eq:B10}
   \end{eqnarray}
   We would like to obtain an inequality as tight as possible, so let us consider the maximization of the RHS of Eq.(\ref{eq:B10}).
   It is maximized when $u \propto F^{-1}Gw$, and the maximal value is $w^T G^{T} F^{-1} G w$.
   We obtain a matrix inequality
   \begin{eqnarray}
      E \ge G^{T} F^{-1} G. \label{eq:B11}
   \end{eqnarray}
   Multiplying by a positive semidefinite matrix $H$ and taking the trace of Eq.(\ref{eq:B11}), we obtain 
   \begin{eqnarray}
      \trace [ H E ] \ge \trace [ H G^{T} F^{-1} G]. \label{eq:B12}
   \end{eqnarray} 
   By substituting $\mathcal{Y} = \mathcal{D}^{N}$, $\theta = \bm{s}$, and $\theta^{\est} = \bm{s}^{\est}_{N}$, we obtain Eq.~(\ref{eq:gCRineq}).   
   When the estimator $\theta^{\est}$ is unbiased, i.e., $\sum_{y\in \mathcal{Y}} p(y|\theta) \theta^{\est}(y) = \theta$, 
   the matrix $G$ is the identity matrix, and we obtain the (standard) Cramer-Rao inequality:
   \begin{eqnarray}
      E \ge F^{-1}.
   \end{eqnarray}

   \section{Conditional Fisher matrices\label{sec:ConditionalFisherMatrices}}
   
   In this section we explain the relation between conditional and unconditional Fisher matrices.
   From a simple calculation, we can obtain
   \begin{eqnarray}
      F_{N} (u^{N}, \bm{s}) = \sum_{D^{N-1}} p(D^{N-1}|\bm{s}) \tilde{F}_{N}(\bm{\Pi}^{N},\bm{s}|D^{N-1}), \label{eq:C1}
   \end{eqnarray}
   where the sum is taken over $D^{N-1}\in\mathcal{D}^{N-1}$.
   This is the reason why $\tilde{F}_{N}$ is called the conditional Fisher matrix of $F_{N}$.
   In statistical parameter estimation theory, it is known that the divergence of the conditional Fisher matrix ($\tilde{F}_{N}\to\infty$ as $N\to\infty$) almost everywhere in $\mathcal{D}^{N}$ is part of a sufficient condition for the convergence (known as \emph{strong consistency} in statistics) of a maximum likelihood estimator \cite{HallHeyde1980}. 
   If we assume that the other elements of the set of sufficient conditions are satisfied, the divergence of the conditional Fisher matrix is sufficient for the convergence of a MLE.
   In this case, from Eq.(\ref{eq:C1}), the unconditional Fisher matrix also diverges ($F_{N} \to \infty$), and this is equivalent to the condition that $\trace [F_{N}^{-1}]\to 0$.
   Therefore, the divergence of the unconditional Fisher matrix is a necessary condition for the convergence of a MLE.
   The divergence of $F_{N}$ is, however, not sufficient for the convergence of a MLE.
   
   We illustrate this with a simple example.
   Suppose that our estimation objective is $\mathcal{O}=\mathcal{S}(\mathbb{C}^2)$.
   At the first trial, we perform a POVM $\bm{\Pi}=\{ \Pi_{\mathrm{T}}, \Pi_{\mathrm{F}}  \}$, where $\Pi_{\mathrm{T}} = \Pi_{\mathrm{F}} = \frac{1}{2}I$. 
   We obtain the outcome $\mathrm{T}$ and $\mathrm{F}$ both with $1/2$ probability.
   When we obtain an outcome $\mathrm{T}$ at the first measurement, we perform standard quantum tomography for the rest of all the trials.
   In this case, a MLE converges to the true state, and the conditional Fisher matrix $\tilde{F}_{N}(u^N |D^{N})$ whose $D^N$ includes $x_{1}=\mathrm{T}$ diverges.
   Let $\tilde{F}_{N}(u^{N} | \mathrm{T})$ denote the conditional Fisher matrix.
   On the other hand, when we obtain $\mathrm{F}$ in the first measurement, we repeat the same POVM $\bm{\Pi}$ for the remaining trials.
   Let $\tilde{F}_{N}(u^N | \mathrm{F})$ denote the conditional Fisher matrix whose $D^N$ includes $x_{1}=\mathrm{F}$.
   In this case, no estimator converges to the true state because the POVM $\bm{\Pi}$ does not give us any information, (the probability distribution is (1/2, 1/2), independent of the true state).
   Then we obtain $\tilde{F}_{N}(u^N | \mathrm{F})=0$.
   The unconditional Fisher matrix is calculated as 
   \begin{eqnarray}
      F_{N}(u^N ,\bm{s}) &=& \frac{1}{2} \tilde{F}_{N}(u^N |\mathrm{T}) + \frac{1}{2} \tilde{F}_{N}(u^N |\mathrm{F}) \\
                                   &=& \frac{1}{2} \tilde{F}_{N}(u^N |\mathrm{T}) \\
                                   &\to& \infty ,
   \end{eqnarray}
   i.e., the unconditional Fisher matrix $F_{N}$ diverges even though no estimator converges to the true state with probability $1/2$.
   Therefore the divergence of $F_{N}$ is necessary, but not sufficient for the convergence of a MLE.
   
   As we can see from the above example, in adaptive experimental designs, the essential characteristic of the scheme is not the unconditional Fisher matrix but the conditional Fisher matrices. 
   In order to make a MLE converge, we need to design an experiment such that almost all (not necessarily strictly all) the conditional Fisher matrices diverge.
   From this point of view, the approximation Eq.(\ref{eq:AoptApprox2}) lies at the heart of adaptive experimental designs.

    \begin{figure*}[htbp]
       \includegraphics[width =0.85\linewidth]{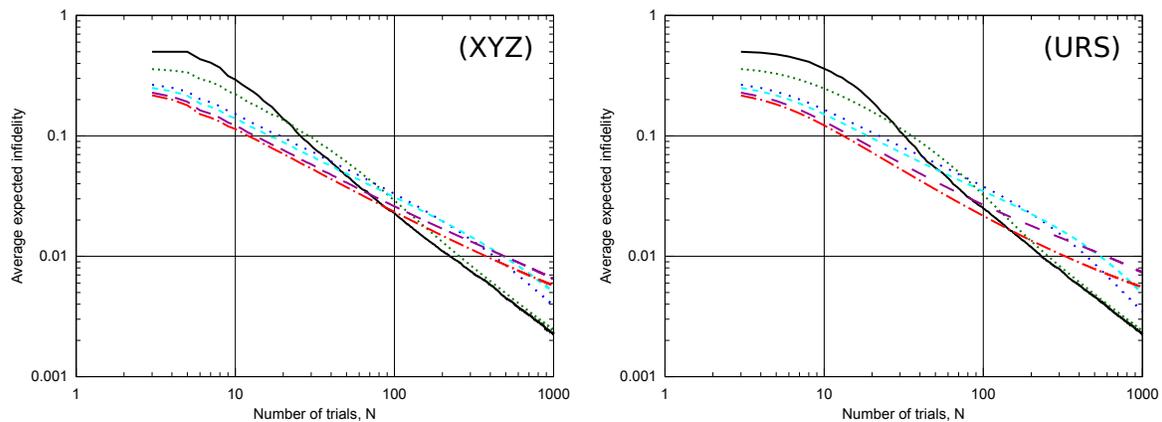}%
       \caption{\label{fig:XYZURS_purity} Purity dependence of average expected infidelity of XYZ and URS schemes: The average is taken over all directions $\theta$ and $\phi$ for each Bloch radius $r$. Average expected infidelity of XYZ repetition (left) and URS (right) for different Bloch radii. Solid line (black): $r=0$, dotted line (green): $r=0.7$, dotted spaced line (blue): $r=0.9$, dashed line (light blue): $r=0.93$, dashed spaced line (purple): $r=0.97$, and dotted dashed line (red): $r=0.99$.
       The number of sequences used for the calculation of the statistical expectation values $N_{\mathrm{mean}}$ is $1000$, and the number of sample points used for the Monte Carlo integration $N_{\mathrm{MC}}$ is $500$ for each Bloch radius $r$.}
   \end{figure*}     
   \section{Purity dependence of XYZ and URS schemes\label{sec:PurityXYZURS}}

      In Fig. \ref{fig:Results_purity} it is shown that the average expected infidelities of XYZ and URS at $N=1000$ have a peak around $r=0.97$.
      Here we explain the origin of the peak.
      Fig. \ref{fig:XYZURS_purity} is a plot of average expected infidelity for six Bloch radii (purities) $r$.
      We choose six purities from the fourteen purities in Fig. \ref{fig:Results_purity} to make things easier to see.
      The average is taken over all directions $\theta$ and $\phi$ for each Bloch radius $r$ ; (XYZ) is for XYZ and (URS) is for URS.
      Roughly speaking, the plots can be interpreted as straight lines with different slopes and $y$-intercepts on a log-log scale.
      As the purity ($r$) increases, two things occur: 
        (i) the slope of the curves becomes less steep, and 
        (ii) the $y$-intercept decreases.
       At $N=1000$, these two effects combine in such a way as to create a peak in the estimation error around $r=0.97$.

%

\end{document}